\documentclass[12pt]{article}
\pdfoutput=1
\usepackage[pdftex]{graphics}
\usepackage{graphicx}
\usepackage{subfigure}
\usepackage{hyperref}
\usepackage{amsmath}
\usepackage[square, comma, numbers, sort&compress]{natbib}

\newcommand\pubnumber{SLAC--PUB--14549}
\newcommand\pubdate{August, 2011}

\def\SLAC{SLAC, Menlo Park, CA 94025}

\def\emailone{\footnote{cbv@stanford.edu}}                     
\def\emailtwo{\footnote{larkoski@stanford.edu}}           

\def\Title#1{\begin{center} {\Large #1 } \end{center}}
\def\Author#1{\begin{center}{ \sc #1} \end{center}}
\def\Address#1{\begin{center}{ \it #1} \end{center}}

\newcommand\pubblock{\rightline{\begin{tabular}{l} \pubnumber\\
         \pubdate \end{tabular}}}
\newenvironment{Abstract}{\begin{quotation} \begin{center}
                       ABSTRACT
     \end{center}\bigskip  }{\end{quotation}}

\def\Acknowledgements{\bigskip  \bigskip \begin{center} \begin{large}
             \bf ACKNOWLEDGEMENTS \end{large}\end{center}}

\def\beq{\begin{equation}}
\def\eeq#1{\label{#1}\end{equation}}
\def\eeqn{\end{equation}}

\newenvironment{Eqnarray}%
   {\arraycolsep 0.14em\begin{eqnarray}}{\end{eqnarray}}
\def\beqa{\begin{Eqnarray}}
\def\eeqa#1{\label{#1}\end{Eqnarray}}
\def\eeqan{\end{Eqnarray}}
\def\CR{\nonumber \\ }

\def\overbar#1{\overline{#1}}

\let\bar=\overbar

\def\lsim{\mathrel{\raise.3ex\hbox{$<$\kern-.75em\lower1ex\hbox{$\sim$}}}}
\def\gsim{\mathrel{\raise.3ex\hbox{$>$\kern-.75em\lower1ex\hbox{$\sim$}}}}

\def\del{\partial}
\def\Dslash{\not{\hbox{\kern-4pt $D$}}}
\def\dslash{\not{\hbox{\kern-2pt $\del$}}}
\def\pslash{\not{\hbox{\kern-2pt $p$}}}
\def\qslash{\not{\hbox{\kern-2pt $q$}}}
\def\kslash{\not{\hbox{\kern-2pt $k$}}}
\def\oneslash{\not{\hbox{\kern-2pt $1$}}}
\def\twoslash{\not{\hbox{\kern-2pt $2$}}}
\def\threeslash{\not{\hbox{\kern-2pt $3$}}}
\def\fourslash{\not{\hbox{\kern-2pt $4$}}}
\def\aslash{\not{\hbox{\kern-2pt $a$}}}
\def\rslash{\not{\hbox{\kern-2pt $r$}}}
\def\Jslash{\not{\hbox{\kern-2pt $J$}}}

\def\msb{{\bar{\scriptsize M \kern -1pt S}}}

\def\drb{{\bar{\scriptsize D \kern -1pt R}}}

\def\spa#1#2{ \langle #1 #2 \rangle }
\def\spb#1#2{ [ #1 #2 ] }
\def\apb#1 {  \langle #1 ] }
\def\bpa#1{  [ #1 \rangle }

\makeatletter
\def\section{\@startsection{section}{0}{\z@}{5.5ex plus .5ex minus
 1.5ex}{2.3ex plus .2ex}{\large\bf}}
\def\subsection{\@startsection{subsection}{1}{\z@}{3.5ex plus .5ex minus
 1.5ex}{1.3ex plus .2ex}{\normalsize\bf}}
\def\subsubsection{\@startsection{subsubsection}{2}{\z@}{-3.5ex plus
-1ex minus  -.2ex}{2.3ex plus .2ex}{\normalsize\sl}}

\renewcommand{\@makecaption}[2]{%
   \vskip 10pt
   \setbox\@tempboxa\hbox{\small #1: #2}
   \ifdim \wd\@tempboxa >\hsize     
       \small #1: #2\par          
     \else                        
       \hbox to\hsize{\hfil\box\@tempboxa\hfil}
   \fi}

 \def\citenum#1{{\def\@cite##1##2{##1}\cite{#1}}}
 
\newcount\@tempcntc
\def\@citex[#1]#2{\if@filesw\immediate\write\@auxout{\string\citation{#2}}\fi
  \@tempcnta\z@\@tempcntb\m@ne\def\@citea{}\@cite{\@for\@citeb:=#2\do
    {\@ifundefined
       {b@\@citeb}{\@citeo\@tempcntb\m@ne\@citea\def\@citea{,}{\bf ?}\@warning
       {Citation `\@citeb' on page \thepage \space undefined}}%
    {\setbox\z@\hbox{\global\@tempcntc0\csname b@\@citeb\endcsname\relax}%
     \ifnum\@tempcntc=\z@ \@citeo\@tempcntb\m@ne
       \@citea\def\@citea{,}\hbox{\csname b@\@citeb\endcsname}%
     \else
      \advance\@tempcntb\@ne
      \ifnum\@tempcntb=\@tempcntc
      \else\advance\@tempcntb\m@ne\@citeo
      \@tempcnta\@tempcntc\@tempcntb\@tempcntc\fi\fi}}\@citeo}{#1}}
\def\@citeo{\ifnum\@tempcnta>\@tempcntb\else\@citea\def\@citea{,}%
  \ifnum\@tempcnta=\@tempcntb\the\@tempcnta\else
  {\advance\@tempcnta\@ne\ifnum\@tempcnta=\@tempcntb \else\def\@citea{--}\fi
    \advance\@tempcnta\m@ne\the\@tempcnta\@citea\the\@tempcntb}\fi\fi}
\makeatother

\def\spa#1#2{\langle #1 #2 \rangle}
\def\spb#1#2{[ #1 #2 ]}
\def\apb#1#2#3{\langle #1  #2  #3 ]}
\def\bpa#1#2#3{[ #1   #2  #3 \rangle}

\begin{document}
\begin{titlepage}
\pubblock

\vfill
\Title{Constructing Amplitudes from Their Soft Limits}
\vfill
\Author{Camille Boucher-Veronneau\emailone and Andrew J. Larkoski\emailtwo}
\Address{\SLAC}
\vfill
\begin{Abstract}
The existence of universal soft limits for gauge-theory and gravity amplitudes
has been known for a long time.  The properties of the soft limits have been exploited
 in numerous ways; in particular for relating an $n$-point amplitude to an $(n-1)$-point amplitude
by removing a soft particle.  Recently, a procedure called inverse soft was developed by 
which ``soft'' particles can be systematically added to an amplitude to construct a higher-point
amplitude for generic kinematics.  We review this procedure and relate it to
Britto-Cachazo-Feng-Witten recursion.  We show that all tree-level amplitudes in gauge theory and
gravity up through seven points can be constructed in this way, as well as certain classes
of NMHV gauge-theory amplitudes with any number of external legs.  This provides us
with a systematic procedure for constructing amplitudes solely from their
soft limits.  
\end{Abstract}
\vfill
\vfill
\end{titlepage}
\def\thefootnote{\fnsymbol{footnote}}
\setcounter{footnote}{1}
%

\section{Introduction}
In the last twenty years, we have seen a resurgence of interest in the S-matrix program of the 60's \citep{analsmatrix} whose goal was to define a quantum field theory through the analytic properties of its S-matrix. The unitarity \citep{Unitarity} and generalized unitarity \citep{GeneralizedUnitarity} methods dramatically simplified loop-level calculations. 
 At tree level, recent interest was triggered in part by Witten's remarkable description of gauge theory as a string theory in twistor space \citep{Witten:2003nn}. Detailed studies of 
an amplitude's analytic properties  have led, in particular, to  the Britto-Cachazo-Feng-Witten (BCFW) recursion relations \citep{BCFref,BCFWref}.  
These relations
exploit the analyticity of the amplitudes in a distinct way from the old S-matrix program:
external particles' momenta are deformed by a complex parameter and the
factorization channels of the deformed amplitude are studied.  This procedure
leads to very compact, on-shell, recursion formulas which have been solved in
generality for ${\cal N}=4$ sYM~\citep{Drummond:2008cr}. 

In the 1980's, it was noticed that photon radiation amplitudes could be expressed as products of soft factors times a lower-point amplitude~\citep{BerendsQED}.
A recent advancement in this area due to Arkani-Hamed, {\it et al.}, 
is a method called {\it inverse soft} which attempts to construct amplitudes from
their soft limits alone \citep{ArkaniHamed:2009dn,nimanotes,twistors3}.  Inverse soft was motivated by studying
the residues of poles in the Grassmannian of \citep{ArkaniHamed:2009dn}.  The Grassmannian
has been conjectured to produce all leading singularities \citep{leadsing} in planar ${\cal N}=4$ sYM from the
residues of its poles.  The relationship between leading singularities at different loop orders, numbers
of external legs and helicity configurations was studied using inverse soft in \citep{Bullimore:2010pa}.
In addition, it has been shown that inverse soft reproduces tree-level maximally-helicity-violating 
(MHV) amplitudes and six-point next-to-MHV (NMHV) amplitudes
in ${\cal N}=4$ sYM \citep{nimanotes}.

In \citep{Nguyen:2009jk}, inverse soft was shown to reproduce 
MHV gravity amplitudes in the form first given in \citep{Bern:1998sv}. Very recently, inspired in part by the inverse-soft procedure discussed in \citep{Nguyen:2009jk}, Hodges presented new expressions for MHV amplitudes \citep{HodgesRecent}.
 However, other than this result, the application of
 inverse soft to gravity amplitudes has been minimal.
This is (possibly) due to several factors: gravity lacks a color expansion (and hence 
color ordering) and the Planck mass is dimensionful.  
Moreover, the gravity soft factor is a gauge-invariant sum of many terms and the inverse-soft procedure exploits each term individually. Thus, there is not a unique form for the gravity soft factor and so there is not a unique procedure for inverse soft in gravity.
Here, we wish to put inverse soft on a firmer ground
for gravity amplitudes, as well as reviewing its application to gauge theory 
amplitudes.  

The main result of this paper is to extend the applicability of inverse soft for gauge theory and 
gravity amplitudes from MHV to NMHV. The gravity soft factor in \citep{Nguyen:2009jk} cannot easily reproduce multiparticle factorization channels necessary to construct NMHV amplitudes. We find that its most natural generalization that can do so is a soft factor inspired by BCFW.  We will use inverse soft to re-express terms in the BCFW expansion 
of an $n$-point 
amplitude $A_n$ as a sum of products of soft factors times a lower point amplitude  $A_{n-m}$. 

For gauge theory we have
\beq
A_n=\sum_{m=1}^{m_{\text{max}}}   \sum_{j=1}^2 \biggl( \prod_{i=1}^m {\cal S'}(p_{i,j}) \biggr)A'_{n-m}(\overbar{p}_{i,j}) \ ,
\eeq{ginvsoft1}
where ${\cal S}$ is a gauge-theory soft factor.
Here, $m$ ranges over the possible BCFW terms; $m=1$ are the two-particle 
factorization terms, $m=2$ the three-particle terms, {\it etc}., and $p_{i,j}$ represent appropriate particles added to the amplitude in the inverse-soft construction. The lower-point amplitude $A_{n-m}(\overbar{p}_{i,j})$ is the $(n-m)$-point amplitude where particles $p_{i,j}$ were removed from $A_n$.
The maximum value of $m$, $m_{\text{max}}$, depends on the helicity configuration of the amplitude and corresponds to the highest-particle factorization channel.  For MHV, $m_{\text{max}}=1$ since there is only two-particle factorization; for NMHV, $m_{\text{max}}$ is at most $n/2 - 1$. The sum over $j$ corresponds to summing over the two distinct $(m+1)$-particle factorization diagrams. In general, to conserve momentum, we 
will also need to deform the individual momenta of particles; this is denoted by primes in Eq.~\ref{ginvsoft1}.
We will show that this formula holds for all MHV amplitudes, for NMHV amplitudes with eight or fewer external legs ($n\leq8$) and for classes of NMHV amplitudes with arbitrary number of legs.

For gravity, we can construct MHV amplitudes as follows:
\beq
M_{\rm MHV}(1,\dotsc,n^{+})=\sum_{i=1}^{n-2}  {\cal G}(n-1,n^{+},i)
M_{\rm MHV}(1,\dotsc,i',\dotsc,(n-1)') \ ,
\eeq{invsoftMHV}
where ${\cal G}(n-1,n^{+},i)$ is a ``gravity soft factor" which is a term in the full gravity soft factor arising when taking particle $n$ soft. As mentioned before, the full gravity soft factor is a sum of many terms and we use each of them individually in our inverse-soft procedure. The sum over $i$ corresponds to summing over the 
$(n-2)$ nonzero two-particle BCFW diagrams or, equivalently, to adding the particle $n$ next to all other possible particles. The primes again denote the deformation of momentum required to make room for the soft particle. At NMHV, we are able to construct amplitudes with seven or 
fewer external particles as
\beq
M_n = \sum_{j=1}^{2(n-2)} {\cal G}(p_j) A'_{n-1}(\overbar{p}_j) + \sum_{j=1}^{(n-2)(n-3)} 
\biggl( \sum_{i=1}^2{\cal G}(p_{i,j})\biggr){\cal G'} (p_j) A'_{n-2}(\overbar{p}_j)\,,
\eeq{invsoft2}
where we sum over the distinct two- and three-particle diagrams in the first and second term respectively. Using our method, four- or higher-particle factorization channels cannot be constructed in gravity which limits us to at most seven-point amplitudes.

We will begin in earnest in Sec.~\ref{softbcfwsection} by studying the two-particle factorization 
terms in the BCFW expansion.  The relationship between these terms and the soft limits of the 
amplitude was first discussed in \citep{PhilipNatalia}.  These terms will be shown to be of the form
${\cal S}(i)A(\overbar{i})$,
where particle $i$ has been removed from the amplitude factor.  We will use this result to uniquely 
define the form of the soft factor that we will use in studying gravity amplitudes.  It should be noted 
that the soft factor we find is distinct from that presented in \citep{Berends:1988zp,Nguyen:2009jk}.  
Since MHV amplitudes only contain two-particle factorization terms in BCFW, we will present a 
compact inverse-soft recursion relation for these amplitudes in gauge theory and gravity.

In Sec.~\ref{multirepro_sec}, we continue by studying higher-point factorization terms in the BCFW 
recursion.  We show explicitly that arbitrary three-particle factorization terms can be built-up from 
two-particle factorization terms using inverse soft.  This leads to the immediate result that any tree-level amplitude in gauge theory and gravity up to seven points can be represented in the form of 
Eq.~\ref{ginvsoft1} or \ref{invsoft2}.  There will be some barriers to constructing arbitrary amplitudes in this manner 
which we will discuss in detail.  However, there exist classes of gauge-theory NMHV amplitudes that can be 
straightforwardly constructed using inverse soft for any number of external legs\footnote{All NMHV gauge-theory
amplitudes can be extracted from a single NMHV ${\cal N}=4$ sYM superamplitude.  One might think that
we could have used a supersymmetric version of inverse soft to construct all NMHV amplitudes.  However, such an approach is not expected to help construct gravity amplitudes as the problem with higher numbers of legs is due to the need to sum over many permutations and there exists only one NMHV graviton amplitude for a given number of legs.}.  We also present 
explicit examples of the procedure for six-point NMHV amplitudes in gauge theory and gravity.

This paper is organized as follows. In Sec.~\ref{gaugegravintro}, we review the soft limits in gauge theory and gravity and 
BCFW recursion. In Sec.~\ref{invsoftsec}, we define inverse
soft precisely and discuss the philosophy of the procedure. As previously discussed, Secs.~\ref{softbcfwsection} and \ref{multirepro_sec} are the meat of the paper where we present the inverse-soft procedure and its relationship to BCFW.
In Sec.~\ref{conclusion}, we present our conclusions. In the appendix, we show the inverse-soft construction of the NMHV six-point gravity amplitude.

\section{Properties of Gauge Theory and Gravity}\label{gaugegravintro}
In this paper, we will consider tree-level gauge-theory and gravity amplitudes
with only gluons and gravitons on the external legs. In a Yang-Mills gauge theory, it is well known 
that tree-level amplitudes can be expanded in a sum of color-ordered partial amplitudes multiplied by
single-trace color factors. For simplicity, we will study these partial amplitudes in the following; the full amplitude can easily be reconstructed from them.  Due to the lack of color ordering, 
amplitudes in gravity contain all 
possible orderings of external legs which will add some complications.

Because all external states are massless, it is very convenient
to work in the spinor-helicity formalism \citep{spinorhelicityref}. 
We will need expressions for the polarization vectors and tensors of gauge theory and gravity
in this formalism.  In gauge theory, because of gauge freedom, the polarization vector
for an external particle $i$ is defined with reference to an arbitrary vector $q$.  Gauge invariant
amplitudes must be independent of $q$, but particularly good choices can simplify computation
greatly.  Explicitly, the polarization vectors are
\beq
\epsilon_+^\mu(i)=\frac{\langle q \gamma^\mu i]}{ \sqrt{2} \spa q i} \ , \qquad 
\epsilon_-^\mu(i)=-\frac{[ q \gamma^\mu i\rangle}{ \sqrt{2} \spb q i}  \ .
\eeq{polarvec}
In gravity, symmetric tensor gauge freedom means that there are two arbitrary vectors $q$ and
$r$ which define the polarization tensor.  The dependence on these reference vectors must be 
symmetrized over and are
\beq
\epsilon_{+}^{\mu\nu}(i)={\langle q \gamma^{(\mu} i] \langle r \gamma^{\nu)} i] \over 2 \spa q i \spa r i} \ , \qquad 
\epsilon_{-}^{\mu\nu}(i)={[ q \gamma^{(\mu} i\rangle [ r \gamma^{\nu)} i\rangle \over 2 \spb q i \spb r i}  \ .
\eeq{polartensor}

With this setup, we can express the soft limits of gauge theory and gravity.
In general, taking an external particle's momentum soft leads to a factorization of the amplitude into a
universal soft factor and a lower-point amplitude where the soft particle has been removed 
\citep{Weinberg:1964ew}.
  The soft factor depends on the momenta
of the particles that were affected by taking that particle soft.  In particular, in a color-ordered
amplitude in gauge theory, only particles adjacent to a soft particle appear in the soft factor.  This
is because only adjacent external particles share a color line.  However, in gravity where there
is no color, all external particles are affected by the limit where one particle goes soft.  All helicity information
of the particle which is taken soft is contained in the soft factor.  The factorization of a
color-ordered amplitude in gauge theory is
\beq
\lim_{j\to 0}A(1,\dotsc,i,j,k,\dotsc,n)=\sqrt{2}\left(- {i\cdot \epsilon(j) \over s_{ij}} + {k\cdot \epsilon(j) \over s_{jk}}\right)
A(1,\dotsc,i,\bar{j},k,\dotsc,n) \ .
\eeq{gaugesoft}
Particles $i$ and $k$ are adjacent to the soft particle $j$ and here $\bar{j}$ 
means that $j$ has been
removed from the amplitude.  The sum in parentheses is the soft factor, is gauge invariant 
and has an especially simple form in spinor-helicity notation.  If $j$ has $+$ helicity, the soft factor is
\beq
\sqrt{2}\left(- {i\cdot \epsilon(j) \over s_{ij}} + {k\cdot \epsilon(j) \over s_{jk}}\right)={\spa i k \over \spa i j \spa j k } \ .
\eeq{gaugesoftplus}

In contrast to gauge theory, there is no such compact form for the soft factor in gravity.  In general the soft 
factor when particle $1$ is taken soft is \citep{Weinberg:1964ew,Berends:1988zp}
\beq
\lim_{1 \to 0}{M(1,2,\dotsc,n) \over M(2,\dotsc,n)}=\sum_{i=2}^n {i^\mu i^\nu \epsilon_{\mu\nu}(1) \over s_{i1}}  \ .
\eeq{gravsoft}
This sum is independent of $q$ and $r$ and using conservation of 
momentum and properties of the
spinor products can be simplified slightly.  Its independence 
on the choice of the reference momenta $q$ and $r$ means that 
the individual terms in the sum can be very different while keeping 
the sum fixed.  In our analysis here,
we will need a particular form for each term and that form will be
 determined to satisfy some simple requirements.  The requirements
 will be discussed in later sections and will be related to the
BCFW on-shell recursion formula, which we now discuss.

The BCFW on-shell recursion relations are an efficient method
for computing amplitudes at tree level in gauge theories  \citep{BCFref,BCFWref} and gravity \citep{gravbcfw}.  
Two external particles, $i$ and $j$, are singled out
and their helicity spinors are deformed as:
\begin{eqnarray}\label{spinordeform}
i\rangle \to i\rangle - z \ j\rangle \ , \qquad i] \to i] \ , \CR
j\rangle \to j \rangle \ , \qquad j] \to j] + z \ i] \ ,
\end{eqnarray}
where $z$ is a complex variable.  The BCFW recursion relation
relates an amplitude to a sum of products of lower-point, on-shell
 amplitudes with momenta of particles $i$
and $j$ deformed as above.  The amplitudes in the sum consist of all possible factorizations
of the amplitude with $i$ and $j$ on opposite sides of the cut.  
These products of amplitudes are evaluated at the value of $z$ 
determined by the location of the pole in the given factorization channel.  If the deformed 
amplitude $A(z)$ vanishes as $z\to \infty$, then the recursion relation is schematically
\beq
A(1,\dotsc,n)=\sum_{\text{R,L}}A_{\text{L}}(\hat{i})\frac{1}{P_{\text{L}}^2}A_{\text{R}}(\hat{j}) \ .
\eeq{BCFW}
Here, the hats indicate to evaluate the amplitude at the shifted momenta and the sum runs over all
possible factorizations.  A BCFW recursion exists in gauge theory and gravity for specific helicity
choices of the deformed particles $i$ and $j$ \citep{jared,gravbcfw}.
  In this paper, we will explicitly develop a relationship between BCFW and 
the soft limits of amplitudes.

In all expressions in this paper, 
we will suppress the gauge and gravity couplings. 
Namely a factor of $i g^{n-2}$ and $i (\kappa / 2)^{n-2}$, where $n$ 
is the number of external particles, is omitted for gauge theory and gravity respectively.   
However, the fact that the gauge coupling is dimensionless
and the Planck mass is dimensionful (in $3+1$ dimensions) 
leads to distinct behavior of the amplitudes in
the soft limits.  

\section{Inverse-Soft Construction of Gauge Theories and Gravity}\label{invsoftsec}

The idea of inverse soft is to ``undo'' the soft limit
of an amplitude.  In particular, in a gauge theory, motivated by Eq.~\ref{gaugesoft}, we can consider the following trial form for an amplitude:
\beq
A(1,\dotsc,i,j,k,\dotsc,n)={\cal S}(i,j,k)A(1,\dotsc,i',\bar{j},k',\dotsc,n) \ .
\eeq{invsoft1}
Here, ${\cal S}$ is a soft factor, for example, that given in Eq.~\ref{gaugesoftplus}.
Unlike in Eq.~\ref{gaugesoft}, no soft limit is taken here and so, to conserve momentum
on the right side, the momenta of adjacent particles $i$ and $k$ must be shifted to compensate for the removal of $j$; this
is indicated by the prime.  The momentum shift depends on the helicity of particle
$j$ and can be expressed as a deformation of the helicity spinors of particles
$i$ and $k$.  If $j$ has $+$ helicity, then the helicity spinors are deformed as
\begin{eqnarray}\label{invsoftdeform}
i\rangle \to i\rangle \ , \qquad i] \to \frac{(i+j)k\rangle}{\spa ik} \ , \CR
k\rangle \to k\rangle \ , \qquad k] \to \frac{(k+j)i\rangle}{\spa ki} \ .
\end{eqnarray}
This deformation conserves momentum: $p_i'+p_k'=p_i+p_j+p_k.$

The expression in Eq.~\ref{invsoft1} only guarantees that the soft limit of $j$ on the 
left side is correct.  To have the correct soft limits for all particles, and so to construct
the amplitude, a sum over a set of
particles $\{j\}$ must be taken on the right.  This sum must also produce all multiparticle
factorization channels present in the true amplitude.
We will show in later sections in specific cases that, by combining several such
inverse-soft terms, the correct multiparticle factorization channels are generated.  This
will be most easily seen in color-ordered gauge theory where only adjacent particles
have nonzero factorization channels.  

Another important point is that the soft factor
${\cal S}$ is unambiguous in gauge theory; it is a single, simple, gauge-invariant
term.  In gravity, the soft factor is a sum of terms,
each of which is not gauge invariant, but the sum is.  Thus, when constructing an object like
Eq.~\ref{invsoft1} in gravity, to even reproduce the soft limit of a single particle, a sum
over particles $i$ and $k$ must be taken.  This opens the possibility of having several
possible forms of inverse-soft construction in gravity.  For each possible form of the terms 
in the gravity soft factor, there exists another possible inverse-soft construction.  

\section{Reproducing the Soft Limits From BCFW Terms}\label{softbcfwsection}
In this section, we will give a precise map from the inverse-soft construction to the BCFW recursion
relations.  In particular, we will show that adding a single particle with the inverse-soft procedure 
is identical to BCFW terms in which one of the amplitude factors is a three-point amplitude.  This
will be straightforward and unambiguous in gauge theory and will be used to define the form of the
soft factor in gravity, as discussed earlier.

First, for a BCFW recursion relation to exist in gauge theory and gravity, the helicity of particles
$i$ and $j$ as in Eq.~\ref{spinordeform} must be $++$, $--$ or $+-$, respectively.
  The $-+$ deformation does not lead to a BCFW recursion and it can be shown 
  that in that case all of the following analysis fails\footnote{This case can be
   studied supersymmetrically, however \citep{nimanotes}.}.  As we are only interested in reproducing 
  the amplitude, we will only need to show that inverse soft can construct all terms in the BCFW
  recursion for a single shift.  In this paper we will only consider the $+-$ shift.  
Also, because only neighboring particles are affected by the soft limit in color-ordered
gauge-theory amplitudes, we will only consider adjacent BCFW deformations.  This is similar
to the deformations in the original discovery of BCF(W) recursion \citep{BCFref}.

\begin{figure}
\centering
    \includegraphics[width=6.3cm]{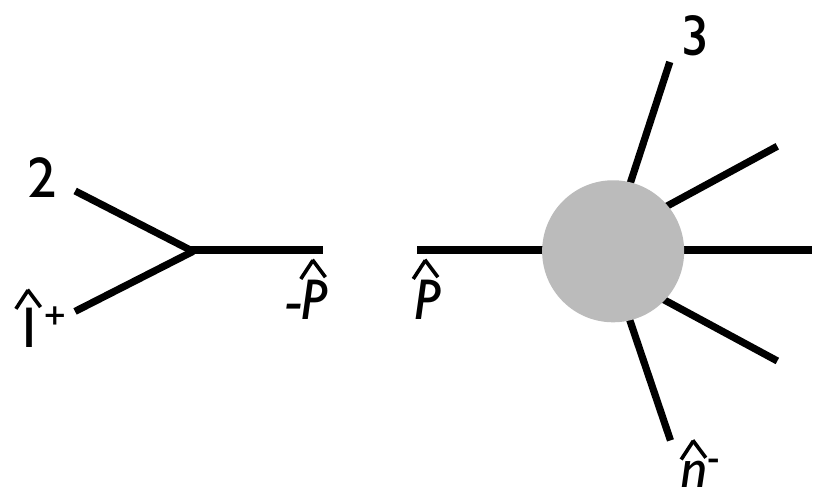}
\caption{Two-particle factorization BCFW diagram. The hatted legs momenta are deformed following Eq.~\ref{mom2part}. }\label{2part}
\end{figure}

The terms in the BCFW recursion we are considering are those whose form is schematically
illustrated in Fig.~\ref{2part}.
   We will refer to these as {\it two-particle factorization channels}, similar to what was used
in \citep{PhilipNatalia}.  This diagram can be expressed as
\beq
D_{+} = A_{\rm L}(\hat{1}^+,2, -\hat{P}) \frac{1}{s_{12}} A_{\rm{R}}(\hat{P},3,\dotsc,\hat{n}^-) \ ,
\eeq{two_bcf}
with the following momentum assignments:
\begin{eqnarray}\label{mom2part}
\hat{1}&=& 1]\langle 1 - {\spa 12 \over  \spa n2} 1]\langle n= {\spa n1 \over \spa n2} 1]\langle 2 \ , \CR
\hat{n}&=& n]\langle n + {\spa 12 \over  \spa n2} 1] \langle n = {(n+1)2\rangle\langle n \over \spa n2} \ , \CR
\hat{P}&=& \hat{1} + 2 ={(2+1)n\rangle\langle 2 \over \spa 2n} \,=\, 2' \ .
\end{eqnarray}
Note that with these assignments, $A_{\rm{R}}$ corresponds to an $n-1$-particle amplitude
with the momenta of particles $n$ and $2$ shifted according to the inverse soft deformation, 
Eq.~\ref{invsoftdeform}.  It now remains to be shown that 
\beq
A_{\rm L}(\hat{1}^+,2, -\hat{P}) \frac{1}{s_{12}}
\eeq{softfact_bcfw}
is the soft factor for particle 1.  We will show this explicitly for either helicity assignment of particle
2 in gauge theory and use the result in gravity to define the soft factor.

\subsection{Gauge Theory} \label{twoPartGauge}

From the two possible helicity assignments of particle 2,
there are two nonzero amplitudes, $A_{\rm L}(\hat{1}^+,2^-, -\hat{P}^+)$ and
$A_{\rm L}(\hat{1}^+,2^+, -\hat{P}^-)$, which evaluate to the same result
when including the shifted momentum.  In particular,
\beq
A_{\rm L}(\hat{1}^+,2^+, -\hat{P}^-)={\spb {\hat 1} 2 ^3\over \spb{2}{({-\hat P})} \spb{({-\hat P})} {\hat 1}}=-{\spb 12 \spa 2n \over \spa 1n } \ .
\eeq{threepoint_gauge}
Including the factor of $1/s_{12}$ gives exactly the soft factor for particle $1$:
\beq
A_{\rm L}(\hat{1}^+,2, -\hat{P}) \frac{1}{s_{12}}={\cal S}(n,1^+,2)=  \frac{\spa n 2  }{\spa n1 \spa 12} \ .
\eeq{soft_bcfw_gauge}
Putting the pieces together, we have shown that two-particle factorized BCFW terms can
be computed with the inverse-soft procedure.  Precisely, in gauge theory,
\beq
A_{\rm L}(\hat{1}^+,2, -\hat{P}) \frac{1}{s_{12}} A_{\rm R}(\hat{P},3,\dotsc,\hat{n}^-)
 = {\cal S}(n,1^+,2) A_{\rm R}(2',3,\dotsc,n')\ ,
\eeq{bcfwinvsoft_eq}
where the hat and prime represent respectively the BCFW (Eq.~\ref{spinordeform}) 
and inverse soft momentum deformations (Eq.~\ref{invsoftdeform}).
With the $P$ and $C$ invariance of pure Yang-Mills theories this result also holds
for a negative helicity soft particle.

For MHV or $\bar{\rm MHV}$ amplitudes in gauge theory, there only exist two-particle factorization channels,
so we can write down an explicit inverse-soft recursion relation for these amplitudes.  
Specializing to MHV, these amplitudes are functions purely of the angle-bracket
spinors and so the momentum shifts of Eq.~\ref{invsoftdeform} do not explicitly appear.  That is,
\beq
A_{\rm MHV}(1,\dotsc,n^+) = {\cal S}(n-1,n^+,1)A_{\rm MHV}(1,\dotsc,n-1)\ ;
\eeq{mhvgaugerec}
to construct an MHV amplitude with one more particle we need only multiply
by the appropriate soft factor.

\subsection{Gravity}\label{twoPartGrav}

Motivated by the relationship between inverse soft and BCFW in gauge theory,
we now turn to considering gravity.  We begin as in gauge theory by constructing
terms in the BCFW recursion relation and then expressing them in inverse-soft language.
Much of the analysis from gauge theory carries over as three-point gravity amplitudes
are just the square of the corresponding gauge-theory amplitudes.  In gravity, we have
\beq
M_{\rm L}(\hat{1}^{+},2^{+}, -\hat{P}^{-})={\spb {\hat 1} 2 ^6\over \spb{2}{({-\hat P})}^2 \spb{({-\hat P})} {\hat 1}^2}={\spb 12 ^2 \spa 2n ^2\over \spa 1n^2 } \ ,
\eeq{threepoint_grav}
where now the helicity labels mean helicity $\pm2$.  Including the propagator factor
as in the gauge theory case gives
\beq
M_{\rm L}(\hat{1}^{+},2, -\hat{P}) \frac{1}{s_{12}}=  \frac{\spa n 2 ^2 \spb 21 }{\spa n1 ^2 \spa 12}\equiv  {\cal G}(n,1^{+},2) \ .
\eeq{soft_bcfw_grav}
We will take this as our definition of the gravity soft factor: an individual term in the sum in
Eq.~\ref{gravsoft}.  To connect with that equation, note that the soft factor here is the $i=2$ term from
Eq.~\ref{gravsoft} with reference momenta $q,r$ of graviton $1$'s polarization tensor
 set equal to the momentum of graviton $n$. Note that in contrast to the gauge-theory soft factor of Eq.~\ref{soft_bcfw_gauge}, Eq.~\ref{soft_bcfw_grav} is not (anti-)symmetric under the exchange of particles~$2$ and $n$. Particle $n$'s momentum is the reference momentum whereas particle $2$ defines the adjacent particle or a term in the sum of Eq.~\ref{gravsoft}. To reproduce the complete soft limit 
 of particle 1 we must sum over terms of the form of Eq.~\ref{soft_bcfw_grav},
replacing $2$ successively by each particle in the amplitude.  We will discuss in the next section
how multiparticle factorization channels are produced with this soft factor.
 
 As in gauge theory, MHV gravity amplitudes are constructed by the BCFW recursion
 from purely two-particle factorizations.  Unlike their gauge-theory counterparts, gravity MHV
 amplitudes have explicit dependence on both angle- and square-bracket spinors.  Unfortunately, this means that a simple inverse-soft recursion relation such as the one that was written down for gauge-theory MHV amplitudes 
 cannot be written down for gravity since the inverse-soft deformations will appear explicitly in the lower-point amplitude.  In any case, the BCFW recursion
 relations imply the inverse-soft recursion relation for MHV gravity amplitudes:
 \beq
 M_{\rm MHV}(1,\dotsc,n^{+})=\sum_{i=1}^{n-2}  {\cal G}(n-1,n^{+},i)
 M_{\rm MHV}(1,\dotsc,i',\dotsc,(n-1)') \ ,
 \eeq{mhvgravrec}
 where the primes on the right side indicate the inverse-soft momentum deformation.
 
 
Nguyen, {\it et al}., introduced in \citep{Nguyen:2009jk} a distinct inverse-soft construction of
gravity MHV amplitudes.  In constructing MHV amplitudes, they used 
\beq
\sum_{i=2}^n {i_\mu i_\nu \epsilon^{\mu\nu}(i) \over s_{i1}}= 
\sum_{i=2}^{n-2} {\spa{i}{n}\spa{i}{(n-1)} [i1] \over \spa{1}{n} \spa{1}{(n-1)} \spa{1}{i}}
\eeq{}
as the form of the soft factor which first appeared in \citep{Berends:1988zp}.  
 Using this soft factor has the benefit of making a larger permutation
symmetry manifest as well as leading to a convenient ``tree formula'' for amplitudes.  This 
tree formula has been shown to reproduce a previously conjectured 
MHV-level result \citep{Bern:1998sv}.  Motivated by this work, Hodges
presented a new formula for MHV gravity amplitudes in \citep{HodgesRecent}.
 In our language, Hodges' formula for MHV amplitudes can be expressed as
 \beq
  M_{\rm MHV}(1,\dotsc,n^{+})=\sum_{i=3}^{n-1}  \frac{[in]\spa1i \spa 2i}{\spa ni \spa 1n \spa 2n}
 M_{\rm MHV}(1',\dotsc,i',\dotsc,(n-1)) \ .
 \eeq{HodgesMHV}
 Note that this equation has one fewer term than our corresponding formula, Eq.~\ref{mhvgravrec}.
 Indeed, these two expressions for MHV gravity amplitudes are related by momentum conservation
 and a Schouten identity.

While the soft factor used in \citep{Nguyen:2009jk,HodgesRecent} leads to nice expressions
for MHV level amplitudes, difficulties arise when continuing to NMHV.
 Factorization channels with
particles $1$ and $2$ on the same side of the factorization have no simple way to be
constructed in this formalism.  As we will see, this is another motivation
 for using the BCFW-inspired soft factor
\beq
\sum_{i=2}^n {i_\mu i_\nu \epsilon^{\mu\nu}(i) \over s_{i1}}= 
\sum_{i=2}^{n-1} {\spa{i}{n}^2 [i1] \over \spa{1}{n}^2 \spa{1}{i}}
\eeq{}
from which three-particle factorization channels can be constructed.  Of course, the two expressions of the
soft factor are equal, but because we work term by term in the sum, the BCFW-inspired soft factor 
leads more directly to multiparticle factorization.

\section{Reproducing NMHV Amplitudes}\label{multirepro_sec}

We will now discuss how to construct NMHV amplitudes using the inverse-soft procedure. Our goal is to express the amplitude as a sum of terms, each of which is a a string of products of deformed soft factors times a lower-point amplitude. Schematically we want
\begin{equation}\label{goalNMHV}
A_{\text{NMHV}} \sim \sum \,\, \Bigg (\prod_{i=1}^{m} {\cal S}'(p_i) \Bigg )A'_{n-m}\,,
\end{equation}
where the $p_i$ have been removed from $A_{n-m}$. The prime indicates that the amplitudes and soft factors are deformed to conserve momentum as particles are removed. We will use the BCFW decomposition to determine which terms to include in the sum so the problem is reduced to expressing all BCFW diagrams entering the amplitude in the form of Eq.~\ref{goalNMHV}. 

 \begin{figure}[t]
\centering
\subfigure[]
{
    \includegraphics[height=3cm]{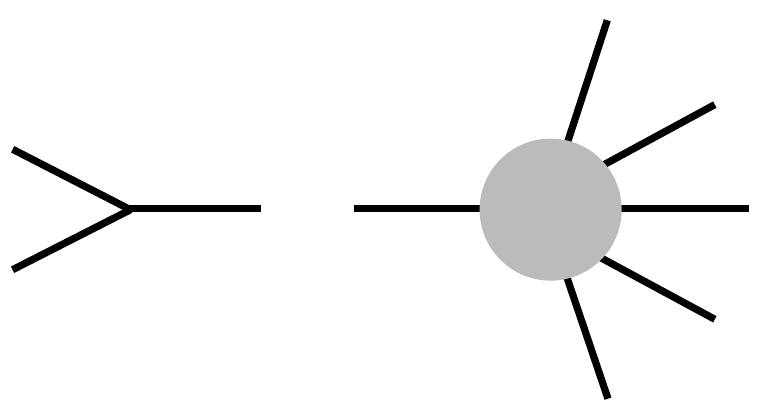}\label{nmhv1}
}
\hspace{.8cm}
\subfigure[]
{
    \includegraphics[height=3cm]{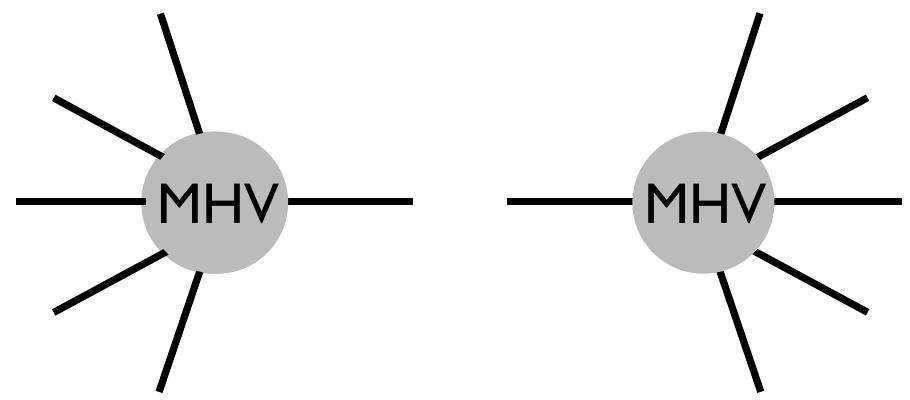}\label{nmhv2}
}
\caption{The two types of diagrams that enter the BCFW decomposition of an NMHV amplitude. 
}
\label{nmhvfact}
\end{figure}

At NMHV level, there are two types of BCFW terms as illustrated in Fig.~\ref{nmhvfact}: those with a three-particle amplitude in the factorization and those with only higher-point MHV amplitudes. The former, illustrated in Fig.~\ref{nmhv1}, was studied in the the previous section where it was expressed as a single soft factor times a deformed amplitude with one less particle.  However, the diagram of Fig.~\ref{nmhv2} is new and we will now discuss how to express it in the form of Eq.~\ref{goalNMHV}. Note that $A_{\rm L}(\hat{1}^+, 2, \ldots, j, -\hat{P})$ is an on-shell MHV amplitude which can be expressed as a product of soft factors times a lower-point amplitude using Eqs.~\ref{mhvgaugerec} or \ref{mhvgravrec}. One approach would then be to express $A_{\rm L}$ for gauge theory this way which would lead to the following expression for the full diagram:
\begin{eqnarray}\label{naiveMulti}
D&=& A_{\rm L}(\hat{1}^+, \ldots, -\hat{P})\, \frac{1}{P^2} \, A_{\rm R} (\hat{P},\ldots,\hat{n}^-) \nonumber \\
&=&{\cal S}(i, j^+, k) \,A_{\rm L}(\hat{1}^+, \ldots,i', \bar{j}, k', \ldots, -\hat{P})\, \frac{1}{P^2} \, A_{\rm R} (\hat{P},\ldots,\hat{n}^-)\,, 
\end{eqnarray}
where particle $j$ has been removed from the amplitude in the second line. While Eq.~\ref{naiveMulti} expresses the diagram as a product of a soft factor times lower-point amplitudes, an extra explicit propagator is present and there is more than one lower-point amplitude. Note also that $\hat{P}$ is the same in both lines of Eq.~\ref{naiveMulti}: it is the sum of all left-hand-side particles' momenta including particle $j$. It is not an original particle whose momentum was deformed using inverse soft (compare with two-particle factorization in Eq.~\ref{mom2part} where $\hat{P}$ is equal to $2'$). Thus, Eq.~\ref{naiveMulti} is not what we want and it does not express the diagram in the desired form of Eq.~\ref{goalNMHV}.

We will then adopt a different approach. We will build up $A_{\rm L}$ times the propagator starting with a three-point amplitude times a propagator factor, 
\begin{equation} \label{nmhvStart}
D_{\text{start}} = A_{\rm L}(\hat{1}^+,j,-\hat{P}){ 1 \over s_{1 j}} A_{\rm R} (\hat{P}, j+1, \ldots,\hat{n}^-) = 
S(n, 1^+, j)  A_{\rm R} (j', j +1,\ldots,n'^-)
\end{equation}
and will add particles until all that were contained in the original $A_{\rm L}$ are included. Namely, we will build the diagram of Fig.~\ref{nmhv2} starting with a two-particle diagram where only two particles are in the left-hand-side MHV amplitude. We add particles in turn to the left amplitude with soft factors and deforming the momenta appropriately. Thus, a three-particle factorization will have two soft factors, a four-particle factorization three and so on.   We do not explicitly add particles to $A_{\rm R}$, but we must make sure that the momentum shifts generated by the work on the left correspond to the correct BCFW deformations. Thus, in order to achieve the form of Eq.~\ref{goalNMHV}, we have two requirements: $A_{\rm L}$ times the propagator must reduce to a product of soft factors and all  inverse-soft shifted momenta must reproduce the BCFW-deformed momenta.

Unfortunately these requirements are not satisfied generically for any diagram of the form illustrated in Fig~\ref{nmhv2}. For instance, we will see that particles' momenta need to be deformed in a specific way and in a specific order to reproduce the BCFW shift. On the other hand, we need to keep $A_{\rm L}$ nonzero throughout the construction which can restrict which particle can be added at a given step. For instance, the nonzero three-point BCFW vertex with particle $\hat{1}^+$ has helicity $(+,+,-)$. Thus, the first particle to be added on the left must have negative helicity. Complications will also arise when trying to add a particle next to the unphysical momentum $\hat{P}$. These caveats will be discussed in detail in the next sections.

In section~\ref{momDeform}, we will discuss how the inverse-soft shift can reproduce the BCFW deformation in $A_{\rm R}$. In sections~\ref{leftGauge} and~\ref{leftGrav}, we will see how $A_{\rm L}$ times the propagator can be constructed as a string of deformed soft factors separately for gauge theory and gravity. Finally, in section~\ref{caveatSection}, we will summarize the above mentioned caveats and address the applicability of the procedure.

\subsection{Momentum Deformation in $A_{\rm R}$} \label{momDeform}

In this section we will show how the inverse-soft shift can be equivalent to the usual BCFW deformation in $A_{\rm R}$ at each step of the construction. We begin with a two-particle factorization diagram in Eq.~\ref{nmhvStart} and the result from Eq.~\ref{mom2part} that the associated deformation is equivalent to the BCFW shift. Here, we will write the deformed momentum differently from Eq.~\ref{mom2part} to make the generalization to higher points easier:
\begin{eqnarray} \label{momShiftStart}
j'&=& 1 + j + \frac{(1+j)^2}{\langle n (1+j) 1]}\,1 ] \langle n\,=\,\hat{P}\,,  \nonumber \\
n'&=&n + \frac{(1+j)^2}{\langle n (1+j) 1]}\, 1]\langle n \,=\, \hat{n}\,.
\end{eqnarray}  
Now, let's add particle $2$ deforming particles $1$ and $j$. Note that since the three-point vertex with $\hat{1}^+$ has helicity $(+,+,-)$, particle $2$ needs to have negative helicity. Thus, the $-$ helicity spinors are shifted and $|1 ]$ is not touched. This leads to $n''$ and $j''$:
\begin{eqnarray} \label{momShift2}
j''&=& 1' + j' + \frac{(1'+j')^2}{\langle n (j'+1') 1]}\,1 ] \langle n  =
1 + 2 + j + \frac{(1+ 2 + j)^2}{\langle n (1+2 +j ) 1]}\,1 ] \langle n = \, \hat{P} \,,
\nonumber \\
n''&=&n + \frac{(1'+j')^2}{\langle n (1'+j') 1]}\, 1]\langle n =
n + \frac{(1+2+j)^2}{\langle n (1+2+j) 1]}\, 1]\langle n  \,=\, \hat{n}\,.
\end{eqnarray}  
We can now add any number of positive helicity particles between particle $2$ and particle $j$. From Eq.~\ref{momShift2}, it is easy to see that the inverse-soft shifts on particles $2,\dotsc,j$ will continue to reproduce the correct BCFW deformation. Namely, at each step of the construction, the momentum assignments in $A_{\rm R}$ corresponds to the momentum assignments of a BCFW diagram with the particles currently present. It is interesting to note that inverse soft gives us a definite procedure to extract the $\hat{P}]$ and $\langle\hat{P}$ components of $\hat{P}$. Using the form of Eq.~\ref{mom2part}, we have:
\begin{equation} \label{PhatSplit}
j'' = {(j'+1')n\rangle \over \spa {j'}n} \langle j' 
=   \frac{[1j](1+2+j)n\rangle}{ \apb {n}{(j+2)}{1}} \times \frac{ [ 1(j+2)}{[1j]} = \hat{P}] \langle \hat{P} \ .
\end{equation}

Finally, note that the particles on the left had to be deformed in a specific order to reproduce the BCFW shift. Namely, in going from the initial two-point diagram to a three-point diagram the two particles on the left need to be deformed (particles $1$ and $j$ in the example above). Afterward, the momentum of particle $1$ can no longer be touched. For instance, particles~$2$ and $j$ would be shifted in Eq.~\ref{momShift2} to add particle~3. Since which momenta get shifted is related to the position of the added particle, this will restrict which diagrams can be constructed. We will discuss this issue in more detail in section~\ref{caveatSection}.

\subsection{A Product of Soft Factors in Gauge Theory} \label{leftGauge}

We now turn to showing that $A_{\rm L}$ times the propagator can be constructed as a string of deformed soft factors. In gauge theory this can be achieved by starting with the three-point amplitude $A_{\rm L}(\hat{1}^+, j, -\hat{P})$ and adding particles between $\hat{1}$ and $j$. This has the nice feature that no explicit $\hat{P}$ appears in the soft factors because particles are never added next to $\hat{P}$. However, note that
\begin{equation}\label{zeroThreePoint}
A_{\rm L}(\hat{1}^+,j^+,-\hat{P}^+) = A_{\rm L}(\hat{1}^+,j^-,-\hat{P}^-) = 0,
\end{equation}
where the second equality follows from the fact that $\langle \hat{P}|$ is proportional to $\langle j|$ as can be seen from Eq.~\ref{mom2part}. One consequence of Eq.~\ref{zeroThreePoint} is that the initial nonzero three-point amplitude will necessarily be  $\overbar{\rm{MHV}}$ which is a function of square brackets. Momentum shifts coming from adding $+$ helicity particles will then be visible. Another more direct consequence is that if we are in the situation of Eq.~\ref{zeroThreePoint}, we will have to start with another three-point amplitude without particle $j$. The latter particle will need to be added through a soft factor which contains $\hat{P}$.

In the case where $A_{\rm L}(\hat{1}^+, j, \hat{P})$ is 
nonzero we begin with the single
 soft factor 
\beq
A(\hat{1}^+,j,-\hat{P}){ 1 \over s_{1 j}} = {\cal S}(n,1^+,j) ,
\eeq{exnmhvsoft}
particles can be added to it.  First add the negative helicity particle, particle $2$ for definiteness, between $1$ and $j$ as
\beq
A(\hat{1}^+,2^-,j,-\hat{P}){ 1 \over s_{12 j}} = {\cal S}(1,2^-,j){\cal S}(n,1'^+,j') \ ,
\eeq{exnmhvsoft1}
where, in the second soft factor, we have to deform the momenta of 
particles $1$ and $j$ according to inverse soft.  Note also that $s_{12j} = s_{1'j'}$. We can now add more positive helicity particles in the exact same way by inserting them one by one between particles $2$ and $j$.  

Consider now the case where both $j$ and $\hat{P}$ have negative helicity. We will start with $A_{\rm L}(\hat{1}^+, 2^+, -\hat{P}^-)$ and add particle $j^-$ between $2$ and $\hat{P}$. We have
\begin{equation}
A_{\rm L}(\hat{1}^+,2^+,j^-,-\hat{P}^-) \frac{1}{s_{12j}} = {\cal S}(2,j^-,- \hat{P}) A_{\rm L}(1^+,2^+, -\hat{P}) \frac{1}{s_{12j}}\,,
\end{equation}
where we used the usual formula, Eq.~\ref{mhvgaugerec}, to decompose the four-point $\overbar{\text{MHV}}$ amplitude and both sides were multiplied by the propagator. Remember that because we are adding a negative helicity particle to a $\overbar{\text{MHV}}$ amplitude, the shifts on $2$ and $\hat{P}$ are not visible in the three-point amplitude. The BCFW shift on particle $1$ also does not appear. We can then instead imagine shifting the angle spinors of particles $1$ and $2$ to make room for particle $j$. Then $1/s_{12j} = 1/s_{1'2'}$. Also note that $P = 1 + 2 + j = 1' + 2'$. Thus, we can write
\begin{eqnarray} \label{gaugePhat}
A_{\rm L}(\hat{1}^+,2^+,j^-,-\hat{P}^-) \frac{1}{s_{12j}}&=&{\cal S}(2,j^-, \hat{P}) A_{\rm L}(1'^+,2'^+, -\hat{P}^-) \frac{1}{s_{1'2'}} \nonumber \\
&=&{\cal S}(2,j^-, \hat{P})\,  {\cal S}(n, 1'^+, 2') \nonumber \\
&=&{\cal S}(2,j^-, 2'')\,  {\cal S}(n, 1'^+, 2')\,,
\end{eqnarray}
where we used Eq.~\ref{soft_bcfw_gauge} to combine the soft factor with the propagator. 
In going from the second line to third line we have replaced $\hat{P}$ by $2''$ as in Eq.~\ref{PhatSplit}.
Note also that the gauge-theory soft factor is symmetric in $\hat{P}\leftrightarrow -\hat{P}$.

It is insightful to do an explicit example. Inserting the soft-factor expressions into Eq.~\ref{gaugePhat}:
\begin{eqnarray}
{\cal S}(2,j^-, 2'') \, {\cal S}(n, 1'^+, 2')&=& \frac{[2 2'']}{[2j][j2'']} \, \frac{\spa{n}{2'}}{\spa{n}{1'}\spa{1'}{2'}} \nonumber \\
&=& \frac{[2 (1+j) n\rangle}{[2j][j(1+2)n\rangle} \, \frac{\apb{n}{(2+j)}{1}[12]}{\apb{n}{(1+j)}{2}(1+2+j)^2} \nonumber \\
&=& \frac{\apb{n}{(2+j)}{1}[12]}{\bpa{j}{(1+2)}{n}[2j](1+2+j)^2} \,,
\end{eqnarray}
where we used the value of $2'']$ given in Eq.~\ref{PhatSplit}. We can now compare this with the expression obtained directly through BCFW,
\begin{eqnarray}
A_{\rm L}(\hat{1}^+,2^+,j^-,-\hat{P}^-) \frac{1}{s_{12j}} &=& \frac{[12]^3}{[2j][j\hat{P}][\hat{P}1]} \, \frac{1}{s_{12j}} \nonumber \\
&=& \frac{\apb{n}{(2+j)}{1}[12]}{\bpa{j}{(1+2)}{n}[2j](1+2+j)^2} \,,
\end{eqnarray}
and it agrees with the inverse-soft construction as expected.

In summary, it is always possible to go from a two-particle to a three-particle diagram using inverse soft. The 
momenta of the two particles initially on the left are always deformed irrespective of the location of the added third 
particle. One can also see that it will not be possible to add particles next to $\hat{P}$ after this first step.  
Each time a particle is added adjacent to $\hat{P}$, the momentum of $1$ must be deformed to conserve
momentum.  In constructing a four-particle factorization channel we add a $+$ helicity particle
which would induce a shift of $|1]$, destroying the equivalence of the BCFW and inverse soft shifts in
$A_{\rm R}$; see Eq.~\ref{momShift2}. For the same reason, it is not possible to add a $+$ helicity particle next to particle~$1$.

\subsection{A Product of Soft Factors in Gravity} \label{leftGrav}

As there is no color ordering in gravity, soft particles must be inserted at all possible locations. This includes inserting the particle
adjacent to the unphysical particle $\hat{P}$. As mentioned above, it is not possible to add next to $\hat{P}$ in a three-particle diagram to create a four-particle diagram.  Consequently, in gravity, at most three-particle factorization diagrams can be created with inverse soft. Adding particle~$2$ next to particles~$j$ and $\hat{P}$ with particle~$1$'s momentum the reference momentum (see Eq.~\ref{soft_bcfw_grav}), we find
\begin{eqnarray}\label{exgravnmhvsoft}
M(\hat{1}^{+},2^{-},j,-\hat{P}){ 1 \over s_{12 j}} &=& \bigl[{\cal G}(1,2^{-},j)+{\cal G}(1,2^{-},-\hat{P})\bigr]{\cal G}(n,1'^{+},j') \\ \nonumber
&=& \bigl[{\cal G}(1,2^{-},j)-{\cal G}(1,2^{-},j'')\bigr]{\cal G}(n,1'^{+},j')\ .
\end{eqnarray}
Note that the gravity soft factor is antisymmetric in $j'' \leftrightarrow -j''$ which
explains the minus sign in the second line of Eq.~\ref{exgravnmhvsoft}.
To prove this, write the explicit values for the soft factors in the equation above:
\begin{eqnarray}\label{ExplicitSoft}
\bigl[{\cal G}(1,2^{-},j)-{\cal G}(1,2^{-},j'')\bigr]{\cal G}(n,1'^{+},j') &=&
\Bigg[ \frac{[j1]^2\spa{2}{j}}{[21]^2[j2]} -\frac{[j''1]^2\spa{2}{j''}}{[21]^2[j''2]} \Bigg ]  \frac{\spa{n}{j'}^2\spb{1}{j}}{\spa{n}{1'}^2\spa{j'}{1'}} \nonumber \\
&=&\frac{[1j]\spa j 2\apb{n}{(1+2)}{j}}{[21][j2]\apb{n}{(j+1)}{2}} \times \frac{\apb{n}{(j+2)}{1}^2[j1]^2}{\apb{n}{(1+2)}{j}^2 P^2} \nonumber \\
&=& \frac{[1j]^3\spa{j}{2}\apb{n}{(j+2)}{1}^2}{[21][j2]\apb{n}{(j+1)}{2}\apb{n}{(1+2)}{j}P^2} \,,
\end{eqnarray}
where we used the explicit value of $j''$ given in Eq.~\ref{PhatSplit}. For completeness, we will compute the left-hand side of Eq.~\ref{exgravnmhvsoft} directly using the BCFW procedure:
\begin{eqnarray}\label{ExplicitBCFW}
M(\hat{1}^{+},2^{-},j,-\hat{P}){ 1 \over s_{12 j}}&=&
-\frac{\spa{2}{j}[\hat{P}1]^6}{[2j][2\hat{P}][21][j\hat{P}][j1]} \times \frac{1}{P^2}
\nonumber \\
&=& \frac{[1j]^3\spa{j}{2}\apb{n}{(j+2)}{1}^2}{[21][j2]\apb{n}{(j+1)}{2}\apb{n}{(1+2)}{j}P^2} \,.
\end{eqnarray}
Thus, the expression built using inverse soft agrees with the usual BCFW construction as expected.

The requirement for a consistent inverse-soft construction, that such
unphysical soft factors be included in constructing multiparticle gravity 
factorization channels, is very restrictive.  The soft factor we have used to define 
the gravity inverse-soft factor is exactly what is needed to reproduce individual BCFW
terms.  As mentioned earlier, other soft factors, such as that from \citep{Nguyen:2009jk},
could have significant difficulty in reproducing the multiparticle factorization channels
precisely because of this issue.

\subsection{Applicability of the procedure} \label{caveatSection}

In this section, we will summarize which diagrams and which NMHV helicity amplitudes can be constructed with the inverse-soft  procedure discussed above. First, for gravity, we can construct all two- and three-particle factorization diagrams. However, as previously mentioned, four-particle diagrams cannot be created as they require adding a particle next to $\hat{P}$ in a three-particle diagram which destroys the BCFW deformation in $A_{\rm R}$. Thus,  only NMHV gravity amplitudes with seven or fewer external particles ($n \le 7$) can be constructed using inverse soft.

In gauge theory, we can also reproduce all two- and three-particle diagrams allowing us to construct all $n \le 7$ amplitudes. However, we can do more as we can add particles between particles~$2$ and $j$ to create four- and higher-point diagrams. Namely, we can create diagrams of the form 
\begin{eqnarray}
&&A_{\rm L}(\hat{1}^+, 2^-, 3^+, \ldots, j^{\pm}, -\hat{P}^{\mp}) \frac{1}{P^2} A_{\rm R}(\hat{P}^{\pm}, \ldots, \hat{n}^-)\,, \\
&&A_{\rm L}(\hat{1}^+, 2^+, 3^+, \ldots, j^{-}, -\hat{P}^-) \frac{1}{P^2} A_{\rm R}(\hat{P}^+, \ldots, \hat{n}^-)\,,
\end{eqnarray}
where in the first line we start with $A_{\rm L}(\hat{1}^+, j^{\pm}, -\hat{P}^{\mp})$, add particle $2^-$, and fill in with positive helicity particles between particles~$2$ and $j$. In the second line, we start with $A_{\rm L}(\hat{1}^+, 2^{+}, -\hat{P}^{-})$, add particle $j^-$ next to $\hat{P}$, and fill in again between $2$ and $j$. 

Note that we can also take $A_{\rm L}$ as our lower point amplitude and add particles to $A_{\rm R}$. This will allow us to construct the additional diagrams
\begin{eqnarray}
&&A_{\rm L}(\hat{1}^+, \ldots, -\hat{P}^{+}) \frac{1}{P^2} A_{\rm R}(\hat{P}^{-}, (j+1)^+, \ldots, (n-1)^+, \hat{n}^-) \,,\\
&&A_{\rm L}(\hat{1}^+, \ldots, -\hat{P}^{-}) \frac{1}{P^2} A_{\rm R}(\hat{P}^{+}, (j+1)^+, \ldots, (n-1)^-, \hat{n}^-)\,,
\end{eqnarray}
where in the first line we start with $A_{\rm R}(\hat{P}^-, (n-1)^+, n^-)$, add particle $(j+1)^+$, and fill in with positive helicity particles between particles~$(j+1)$ and $(n-1)$. In the second line, we start with $A_{\rm R}(\hat{P}^+, (n-1)^-, n^-)$, add particle $(j+1)^+$ next to $\hat{P}$, and fill in again between $(j+1)$ and $(n-1)$. 

\begin{figure}
\centering
    \includegraphics[width=7.5cm]{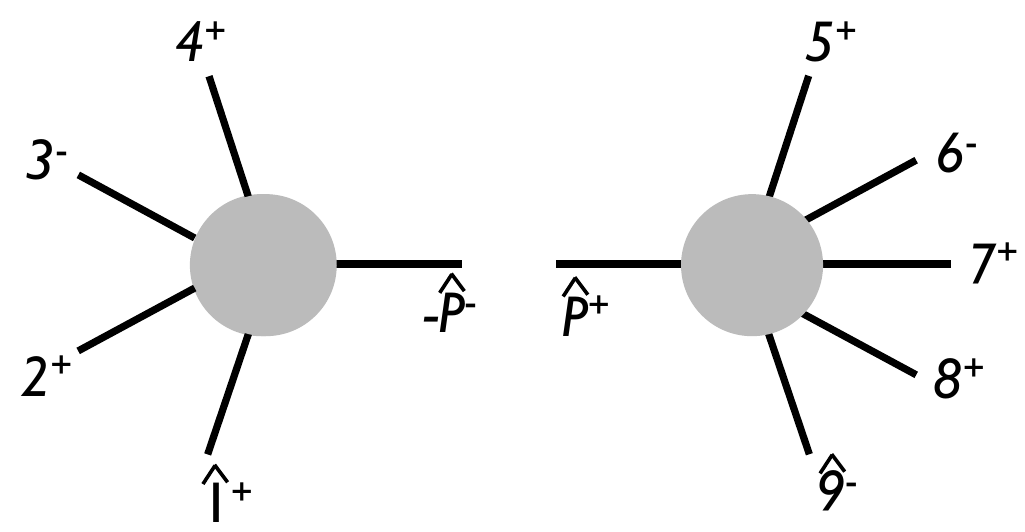}
\caption{A BCFW diagram for the amplitude $A(1^-,2^+,3^-,4^+,5^+,6^-,7^+,8^+,9^-)$
which cannot be constructed with inverse soft.}\label{9ptfail}
\end{figure}

We now notice that any NMHV amplitude containing adjacent particles  
with helicity $(-,+,-)$ or $(-,-,+)$ can be created using the inverse-soft procedure we have 
developed here.  In such a case, we factorize the amplitude in BCFW at the location of the 
$+$ helicity particle in this series: either $(-,\hat{+}|\hat{-})$ or $(-,\hat{-}|\hat{+})$.  This 
factorization results in the possible constructible terms discussed above.  Such a factorization
is always possible in NMHV gauge-theory amplitudes up through eight external legs.  Problems begin
at nine points as illustrated in Fig.~\ref{9ptfail}.  The amplitude 
$A(1^+,2^+,3^-,4^+,5^+,6^-,7^+,8^+,9^-)$ does not contain either of the strings of adjacent
particles with helicity $(-,+,-)$ or $(-,-,+)$.  The BCFW diagram in 
Fig.~\ref{9ptfail} would require adding two particles adjacent to $\hat{P}$ or adding a $+$
helicity particle next to particle~$1$ and thus cannot be constructed using our inverse-soft procedure.

\subsection{Example: Gauge Theory NMHV}\label{gaugeexamplesec}
To show the utility of inverse soft, we will give an explicit example of constructing the 
six-point NMHV gauge-theory amplitude using inverse soft. 
Consider the gauge-theory amplitude $A(1^+,2^-,3^+,4^-,5^+,6^-)$. 
 We will deform the momenta of particles $3$ and $4$ for the BCFW construction.  This deformation leads to 
 three factorizations which contribute to the amplitude:
\begin{eqnarray}\label{BCFWbreak}
D_1&=&A(2^-,\hat{3}^+,-\hat{P}) \frac{1}{s_{23}}  A(\hat{P},\hat{4}^-,5^+,6^-,1^+) \ ,\\ \nonumber
D_2&=&A(1^+,2^-,\hat{3}^+,-\hat{P}) \frac{1}{s_{123}}  A(\hat{P},\hat{4}^-,5^+,6^-) \ ,\\ \nonumber
D_3 &=& A(6^-,1^+,2^-,\hat{3}^+,-\hat{P}) \frac{1}{s_{45}}  A(\hat{P},\hat{4}^-,5^+) \ .
\end{eqnarray}
We will consider each of these terms in turn and show that they can be written simply in
the inverse-soft construction language. 
For compactness, we will only decompose these diagrams down to soft factors times MHV or $\overbar{\text{MHV}}$ amplitudes.
  
We first consider $D_1$.  As shown earlier, for the choice of helicity of $P$ that gives
a nonzero value, $A(2^-,\hat{3}^+,-\hat{P}) (1/s_{23}) = {\cal S}(2,3^+,4).$ Thus, we can 
equally express $D_1$ as
\beq
D_1 =  {\cal S}(2,3^+,4) A (2'^{-},1^+,6^-,5^+,4'^-)\ .
\eeq{D1}
Note that the amplitude that remains is of $\bar{\text{MHV}}$-type and so the inverse soft
momentum shift explicitly appears.  The same arguments hold for $D_3$; this time, re-expressing
the right side of the factorization.  $D_3$ can be written as
\beq
D_3 = {\cal S }(3,4^-,5)A(6^-,1^+,2^-,3'^+,5'^+) \ .
\eeq{D3}

$D_2$ is slightly more complicated and requires some care.  We will rewrite the left side of the
BCFW factorization as a product of two soft factors as in Eq.~\ref{exnmhvsoft1}.  
To do this, we will start with the soft
factor $S(1,3^+,4)$ and add particle $2$ between $1$ and $3$.  This gives
\beq
A(1^+,2^-,\hat{3}^+,-\hat{P}) \frac{1}{s_{123}}={\spb 13 \over \spb 12 \spb 23}
 { \spa {1'}4 \over \spa {1'}{3'} \spa{3'}4}\equiv {\cal S}(1,2^-,3){\cal S}(1',3'^+,4) \ ,
\eeq{2bt13}
where the primes indicate the inverse soft momentum deformation with 
negative helicity particle $2$.  The right side of the factorization is subtle but proceeds
exactly as previously.  We first
add particle $3$ to the expression.  This produces $A(1'^+,6^-,5^+,4'^-)$, where
the primes indicate the inverse soft momentum deformation with particle $3$.  Next, we add 
particle $2$ between particles $1$ and $3$.  Thus, we now only deform the momentum of 
particles $1$ and $3$ by the momentum of $2$.  
To see how this affects the momentum of
the particles in $A(1'^+,6^-,5^+,4'^-)$, consider first $1'$.  We have
\beq
1'= { (1+3) 4\rangle \over \spa 14} \langle 1 \ .
\eeq{oneprime}
Deforming the momentum of $1$ and $3$ appropriately gives $1''$:
\beq
1''= {(1+2+3)4\rangle [ 3 (1+2) \over \apb{4}{(1+2)}{3}} \ . 
\eeq{onepprime}
As it must, this equals the momentum flowing through the BCFW cut, $\hat{P} = 1+2+\hat{3}$.
We can now consider the deformation of $4'$.  Starting with
\beq
4'={(4+3)1\rangle \over \spa 41}\langle 4 \ ,
\eeq{fourprime}
$4''$ is then
\beq
4''=4]\langle 4 + {(1+2+3)^2 \over \apb{4}{(1+2)}{3}} 3]\langle 4 = \hat{4} \ .
\eeq{fpprime}
Finally, we can express $D_2$ in inverse-soft language as
\beq
D_2={\cal S}(1,2^-,3){\cal S}(1',3'^+,4) A(1''^+,6^-,5^+,4''^-) \ ,
\eeq{D2}
where the primes and double primes are as above.

Putting it all together, we can express the six-point, alternating helicity, NMHV amplitude
in gauge theory as
\begin{eqnarray}\label{ampinvsoft}
A(1^+,2^-,3^+,4^-,5^+,6^-) &=& {\cal S}(2,3^+,4) A (2'^{-},1^+,6^-,5^+,4'^-) \CR
&& \!\!\!\!\!\!\!+  \  {\cal S }(3,4^-,5)A(6^-,1^+,2^-,3'^+,5'^+) \CR
 && \!\!\!\!\!\!\! +  \  {\cal S}(1,2^-,3){\cal S}(1',3'^+,4) A(1''^+,6^-,5^+,4''^-) \ .
\end{eqnarray}
We present the inverse-soft expression of the NMHV six-point gravity amplitude
in the appendix.

\section{Conclusions}\label{conclusion}

We have shown that inverse soft, with BCFW as our guide, can be used to construct
gauge-theory and gravity amplitudes.  In particular, for specific amplitudes, each term in the BCFW expansion can be built up
from soft factors multiplied by a lower-point amplitude.  This procedure works for all tree-level
gauge-theory and gravity amplitudes with seven or fewer external legs because, as was shown in 
Sec.~\ref{multirepro_sec}, inverse soft can produce three-particle factorization BCFW diagrams.
Also, certain classes of NMHV gauge-theory amplitudes can be constructed with any number of legs
which contain a set of consecutive particles with helicities $(-,-,+)$ or $(-,+,-)$.  

As we have developed it, inverse soft does not explicitly need information from the collinear limits
of amplitudes to be able to reconstruct the amplitude.  Presumably, inverse soft could be used to compute
amplitudes in any massless theory with universal soft limits and a BCFW recursion.
  This property may seem surprising, especially
because inverse soft can be used to construct NMHV amplitudes.  However, this property is similar to
the way that BCFW exploits {\it complex} factorization to construct amplitudes.  In that case, only a subset
of factorization channels are needed to reproduce all factorization channels present in an amplitude.
One can imagine constructing an inverse soft/collinear procedure which incorporates information about
both the soft and collinear limits to create amplitudes.  Perhaps such a procedure is necessary
to extend the applicability to arbitrary helicity configurations and number of legs. 

Nevertheless, it is interesting to consider how far inverse soft can be extended.  In particular, it should be
possible to use inverse soft to construct higher-point loop amplitudes, perhaps along the lines of 
\citep{Bullimore:2010pa}.  Using the supersymmetric BCFW formalism of \citep{ArkaniHamed:2008gz},
supersymmetric inverse soft could be used to construct arbitrary NMHV gauge-theory amplitudes.  This is because 
we have an inverse-soft construction valid for classes of NMHV amplitudes with any number of legs and the 
various helicity orderings of NMHV Yang-Mills amplitudes can be extracted from a unique ${\cal N}=4$ sYM NMHV 
superamplitude. However, supersymmetry would likely not be useful in finding a gravity all-point NMHV inverse
soft construction because the problems we encountered were caused by the need to sum over many permutations. 
  To go further, it is
unclear if BCFW should continue to be our guide.  Indeed, in gravity, it is not clear what the ``best'' or most
useful form of the gravity soft factor is for inverse soft.

Inverse soft has a particularly simple form in the twistor-space representation \citep{twistors3}.  Nguyen, {\it et al.}, showed that their tree formula for gravity
MHV amplitudes also has a twistor-space representation.   In \citep{HodgesRecent}, Hodges conjectured a new BCFW recursion relation in ${\cal N} = 7$ supertwistor space and used it to construct an inverse-soft procedure for MHV amplitudes. It would then be interesting to see if ${\cal N} = 7$ BCFW could be used as a guide to construct multiparticle factorization channels and NMHV amplitudes with a possibly different soft factor. 
It would also be interesting to study whether or not our inverse-soft procedure for
gravity extended to ${\cal N}=8$ supergravity has a twistor-space representation.  
If so, what is the representation?  Are there symmetries of ${\cal N}=8$ amplitudes
that become apparent, analogous to dual superconformal symmetry \citep{Drummond:2008vq}
 and the Yangian
in ${\cal N}=4$ sYM \citep{Drummond:2009fd}?  How does the $E_{7(7)}$ symmetry of 
the moduli space \citep{Cremmer:1979up} manifest itself?  More work
in these directions is necessary to fully elucidate the symmetries and simplicity of scattering amplitudes.

\Acknowledgements
The authors thank Jared Kaplan for initiating our interest and very helpful discussions. We also thank Lance Dixon and Michael Peskin for very useful comments on the manuscript.
This work is supported by the US Department of Energy under 
contract DE--AC02--76SF00515.  A.L.~is also supported by
an LHC Theory Initiative Travel Award. C.B.V.~is supported in part by a postgraduate scholarship from the Natural Sciences and Engineering Research Council of Canada.

\clearpage

\appendix

\section{Six-point Gravity NMHV Amplitude}
In this appendix, we present the six-point NMHV gravity amplitude as 
expressed using inverse soft.  We will consider the amplitude $M(1^+,2^+,3^+,4^-,5^-,6^-)$
and its BCFW representation by deforming the momenta of particles $3$ and $4$.  There are
13 BCFW diagrams and using the techniques of Sec.~\ref{multirepro_sec} can be expressed
as:
\begin{eqnarray}\label{grav6ptNMHV}
M(1^+,2^+,3^+,4^-,5^-,6^-)&=& {\cal G}(4,3^+,1)M(1'^+,2^+,4'^-,5^-,6^-) \CR
&&+ \  {\cal G}(4,3^+,2)M(1^+,2'^+,4'^-,5^-,6^-) \CR
&&+ \ {\cal G}(4,3^+,5)M(1^+,2^+,4'^-,5'^-,6^-) \CR
&&+ \ {\cal G}(4,3^+,6)M(1^+,2^+,4'^-,5^-,6'^-) \CR
&&+ \ {\cal G}(3,4^-,1)M(1'^+,2^+,3'^+,5^-,6^-) \CR
&&+ \ {\cal G}(3,4^-,2)M(1^+,2'^+,3'^+,5^-,6^-) \CR
&&+ \ {\cal G}(3,4^-,5)M(1^+,2^+,3'^+,5'^-,6^-) \CR
&&+ \ {\cal G}(3,4^-,6)M(1^+,2^+,3'^+,5^-,6'^-) \CR
&&+ \ \bigl[{\cal G}(3,5^{-},1)-{\cal G}(3,5^{-},1'')\bigr]{\cal G}(4,3'^{+},1')M(1''^+,2^+,4''^-,6^-) \CR
&&+ \ \bigl[{\cal G}(3,5^{-},2)-{\cal G}(3,5^{-},2'')\bigr]{\cal G}(4,3'^{+},2')M(1^+,2''^+,4''^-,6^-) \CR
&&+ \ \bigl[{\cal G}(3,6^{-},1)-{\cal G}(3,6^{-},1'')\bigr]{\cal G}(4,3'^{+},1')M(1''^+,2^+,4''^-,5^-) \CR
&&+ \ \bigl[{\cal G}(3,6^{-},2)-{\cal G}(3,6^{-},2'')\bigr]{\cal G}(4,3'^{+},2')M(1^+,2''^+,4''^-,5^-) \CR
&&+ \  \bigl[{\cal G}(3,6^{-},5)-{\cal G}(3,6^{-},5'')\bigr]{\cal G}(4,3'^{+},5')M(1^+,2^+,4''^-,5''^-)\ .\CR
\end{eqnarray}
As in Sec.~\ref{momDeform}, primes indicate inverse soft momentum deformations with
 respect to the last added particle (corresponding to the leftmost soft factor) 
 and double primes indicate inverse soft momentum 
 deformations with respect to the first added particle and then the second added particle.

\end{document}